%% file: arxiv.tex
\documentclass{llncs}

\usepackage{amsmath,amssymb,amsfonts}
\usepackage{theorem}
\usepackage{pstricks,pst-node}
\usepackage{listings}

\theorembodyfont{\upshape}

\newcommand{\comment}[1]{}

\newcommand{\set}[1]{ \{ #1 \} }

\newcommand{\band}{\wedge}
\newcommand{\bor}{\vee}
\newcommand{\bnot}{\neg}
\newcommand{\Bimplies}{\Rightarrow}
\newcommand{\vtrue}{\mathsf{true}}

\title{$k$-Step Relative Inductive Generalization}
\author{Aaron R. Bradley}
\institute{Dept. of Electrical, Computer \& Energy Engineering \\
University of Colorado at Boulder \\
Boulder, CO 80309 \\
{\tt bradleya@colorado.edu}
}
\date{\today}

\begin{document}

\maketitle
\begin{abstract}
We introduce a new form of SAT-based symbolic model checking.  One
common idea in SAT-based symbolic model checking is to generate new
clauses from states that can lead to property violations.  Our
previous work suggests applying induction to generalize from such
states.  While effective on some benchmarks, the main problem with
inductive generalization is that not all such states can be
inductively generalized at a given time in the analysis, resulting in
long searches for generalizable states on some benchmarks.  This paper
introduces the idea of inductively generalizing states relative to
$k$-step over-approximations: a given state is inductively generalized
relative to the latest $k$-step over-approximation relative to which
the negation of the state is itself inductive.  This idea motivates an
algorithm that inductively generalizes a given state at the highest
level $k$ so far examined, possibly by generating more than one
mutually $k$-step relative inductive clause.  We present experimental
evidence that the algorithm is effective in practice.
\end{abstract}

\newcommand{\vx}{\bar{x}}
\newcommand{\notmodels}[2]{#2 \not\models #1}
\newcommand{\mymodels}[2]{#2 \models #1}

\section{Introduction}

Several themes for SAT-based symbolic model checking
\cite{Burch+Others/1990} have been explored over the past decade
\cite{Biere+Others/1999,Sheeran+Others/2000,McMillan/2002,McMillan/2003,DeMoura+Others/2003,Bradley+Manna/2007}.
A subset of these methods
\cite{McMillan/2002,DeMoura+Others/2003,Bradley+Manna/2007} derive new
search-constraining clauses from discovered states that lead to
property violations.  In previous work, we introduced induction as one
means of generalizing from such states.  Given a cube $c$ that one
would like to exclude because the states that it describes lead to
violations of a desired property, a \emph{minimal inductive subclause}
$d$ of $\bnot c$ is a clause whose literals are negations of those
appearing in $c$ ($d \subseteq \bnot c$) and that is inductive
relative to known reachability information \cite{Bradley+Manna/2007}.
Not all cubes can be inductively generalized at a given time during
proof construction, however.  This inability to inductively generalize
any given cube (whose satisfying states lead to property violations)
limits the applicability of the technique as previously developed
\cite{Bradley+Manna/2007}: on some benchmarks, the model checker
becomes embroiled in long fruitless searches for generalizable cubes.
However, its success on some nontrivial benchmarks indicates that the
fundamental idea of inductive generalization from states is worth
exploring \cite{Bradley/2007}.

We describe in this paper a method based on induction for generalizing
all cubes (unless the asserted property does not hold).  The algorithm
maintains a sequence $F_{0}, F_{1}, F_{2}, \ldots, F_{k}$ of
over-approximations of sets of states reachable in at most $0, 1, 2,
\ldots, k$ steps, for increasing $k$.  It iteratively generalizes
cubes: a cube $s$ that implies $F_k$ and that leads in one step to
violating the property is inductively generalized relative to the most
general over-approximation $F_{i}$ relative to which the negation of
the state, $\bnot s$, is itself inductive.  If $i<k$, predecessors of
$s$ are treated recursively until $s$ can be inductively generalized
relative to $F_{k}$.  We call this process {\it $k$-step relative
  inductive generalization}.  Once $F_{k}$ is strengthened to the
point that no $F_{k}$-state can transition into a property-violating
state, $k$ is incremented and the generated clauses are propagated
forward through $F_0,F_1,F_2,\ldots,F_{k+1}$ via implication checks.
The iterations continue until convergence (if the property is
invariant) or until discovery of a counterexample trace (if the
property is not invariant).  Section \ref{sec:main} presents this
algorithm in detail.

The symbolic model checker based on $k$-step relative inductive
generalization is robust.  Section \ref{sec:exp} details our
implementation and experiments on the HWMCC 2008 benchmarks
\cite{hwmcc08}.  Our symbolic model checker outperforms the winner of
the {\it unsat} division and the overall winner of the competition.

\section{Preliminaries}
\label{sec:prelims}

\subsection{Definitions}

A \emph{finite-state transition system} $S: (\vx, I, T)$ is described
by a pair of propositional logic formulas: an initial condition
$I(\vx)$ and a transition relation $T(\vx,\vx')$ over a set of Boolean
variables $\vx$ and their next-state primed forms $\vx'$
\cite{Clarke+Others:MC:2000}.  Applying prime to a formula, $F'$, is
the same as priming all of its variables.

A state of the system is an assignment of Boolean values to all $\vx$
and is described by a \emph{cube} over $\vx$, which is a conjunction
of literals, each \emph{literal} a variable or its negation.  The
negation of a cube is a \emph{clause}.  An assignment $s$ to all
variables of a formula $F$ either satisfies the formula, denoted
$\mymodels{F}{s}$, or falsifies it, denoted $\notmodels{F}{s}$.  A
formula $F$ \emph{implies} another formula $G$, written $F \Bimplies
G$, if every satisfying assignment of $F$ satisfies $G$.

A \emph{trace} $s_0,s_1,s_2,\ldots$ of a transition system $S$, which
may be finite or infinite in length, is a sequence of states such that
$\mymodels{I}{s_0}$ and for each adjacent pair $(s_i,s_{i+1})$ in the
sequence, $\mymodels{T}{s_i \band s_{i+1}'}$.  That is, a trace is the
sequence of assignments in an execution of the transition system.  A
state that appears in some trace of the system is \emph{reachable}.

A safety property $P(\vx)$ asserts that only $P$-states (states
satisfying $P$) are reachable.  $P$ is \emph{invariant} for the system
if indeed only $P$-states are reachable.  If $P$ is not invariant,
then there exists a finite \emph{counterexample} trace
$s_0,s_1,\ldots,s_k$ such that $\notmodels{P}{s_k}$.

An \emph{inductive} assertion $F(\vx)$ describes a set of states that
(1) includes all initial states: $I \Bimplies F$, and that (2) is
closed under the transition relation: $F \band T \Bimplies F'$.  An
assertion $F$ is inductive \emph{relative to} another assertion $G$ if
instead of (2), we have that $G \band F \band T \Bimplies F'$.

An inductive \emph{strengthening} of a safety property $P$ is a
formula $F$ such that $F \band P$ is inductive.  Since $F \band P
\Bimplies P$, $F$ is a proof of $P$'s invariance.

\subsection{Inductive Generalization}
\label{subsec:ig}

In previous work, we introduced a technique for discovering a
\emph{minimal inductive subclause} $d$ of a given clause $c$ if one
exists \cite{Bradley+Manna/2007}.  Such a clause $d$ (1) consists only
of literals of $c$ ($d \subseteq c$), (2) is inductive (possibly
relative to known reachability information), and (3) is minimal in
that it does not contain any strict subclauses that are also
inductive.

\emph{Inductive generalization} of a cube $s$ is the process of
finding a minimal inductive subclause $d$ of $\bnot s$, if one exists.
The resulting subclause (if one exists) over-approximates the set of
reachable states while excluding $s$.  In practice, a minimal
inductive subclause is typically substantially smaller than the cube
$s$ from which it is extracted.  Hence, it excludes many other states
as well, which is why we say that the inductive subclause generalizes
that $s$ is unreachable.

\section{Algorithm and Analysis}
\label{sec:main}

We describe a complete symbolic model checking algorithm for safety
properties.  Given a transition system $S: (\vx, I, T)$ and safety
property $P$, it either generates a formula $F$ such that $F \band P$
is inductive or it discovers a counterexample trace.

Section \ref{subsec:informal} presents the algorithm informally, while
Section \ref{subsec:illustration} provides an example of its
application.  Then Section \ref{subsec:formal} formally describes and
proves the correctness of the algorithm.

\subsection{Informal Description}
\label{subsec:informal}

The algorithm constructs a sequence $F_{0}, F_{1}, F_{2}, \ldots$ of
over-approximations of the state sets reachable in at most $0, 1, 2,
\ldots$ steps.  It incrementally refines the sequence until some $F_i$
converges to an inductive strengthening of $P$, or until it encounters
a counterexample trace.

Initially, $F_0 = I$, and $F_i = P$ for $i > 0$, corresponding to the
assumption that $P$ is invariant.  Let $k$ be the level of $F_k$, the
frontier of the sequence.  The sequence satisfies the following
invariants: (1) $F_{0} = I$, (2) $\forall\ 0 \le i < k,\ F_{i}
\Bimplies F_{i+1}$, and (3) $\forall\ 0 \le i < k,\ F_{i} \band T
\Bimplies F_{i+1}'$.  If $F_k \band T \Bimplies P'$, then $F_{k+1}$
becomes the new frontier.  Otherwise, there is a state $s$ that leads
in one step to a violation of $P$.

Given such a state $s$, the algorithm finds the highest level $0 \le i
\le k$ such that $\bnot s$ is inductive relative to $F_{i}$.  If $P$
is invariant, such a level exists.  At this level, $s$ can be
inductively generalized relative to $F_{i}$.

Inductive generalization produces a clause $c \subseteq \bnot s$ that
is inductive relative to $F_{i}$.  It asserts that $s$ --- and any
other state $t$ such that $\notmodels{c}{t}$ --- is not reachable
within $i+1$ steps.  Because $\bnot s$ has been generalized to $c$,
$c$ may exclude states that were previously admitted by some $F_j$ for
$j \le i+1$.  In other words, $c$ potentially represents new $j$-step
reachability information at every level $j$ up to $i+1$.  Therefore,
each $F_j$, for $1 \le j \le i+1$, is strengthened to $F_j \band c$.

If $i = k$, then $s$ has been inductively generalized at the highest
possible level, and $F_k$ no longer admits the state $s$, bringing the
algorithm one step closer to strengthening $F_k$ such that $F_k \band
T \Bimplies P'$.

If $i < k$, then the generalization of $s$ at level $i$ must be pushed
to level $k$.  There must exist some predecessor $p$ of $s$ admitted
by $F_{i+1}$ but excluded by $F_{i}$.  This predecessor is one of the
reasons that $\bnot s$ is not inductive relative to $F_{i+1}$.  Now
$p$ is considered recursively for inductive generalization.  This
recursion continues until $s$ can be inductively generalized relative
to $F_k$.

Once $F_k \band T \Bimplies P'$ holds, the clauses that have been
generated so far are propagated forward through
$F_0,F_1,F_2,\ldots,F_k$: for each clause $d \in \mbox{\tt
  clauses}(F_i)$, if $F_i \band d \band T \Bimplies d'$, then $d$ is
conjoined to $F_{i+1}$.  If the clause sets of two adjacent levels,
$F_i$ and $F_{i+1}$, become equal, then $F_i$ is an inductive
strengthening of $P$ that proves $P$'s invariance.

If $P$ is not invariant, the algorithm discovers a counterexample
trace, though not necessarily a shortest.  Let $s_0,s_1,\ldots,s_n$ be
a shortest counterexample trace.  The algorithm finds a counterexample
trace when $k = n$, if not earlier.  For when $k = n$, each $s_i$, for
$2 \le i \le n$, can be shown to be inductive relative to at most
$F_{i-2}$.  Hence, $s_1$ (or another 1-step state from another
counterexample trace) must eventually be analyzed during the recursion
associated with inductively strengthening $s_n$ (or another state from
another counterexample trace) relative to $F_n$, at which point it
would be found to be reachable from an initial state.

\subsection{An Illustrative Example}
\label{subsec:illustration}

Consider the contrived transition system $S: (\vx, I, T)$ with
variables $\vx = \set{x_0, x_1, x, y_0, y_1, y, z}$, initial condition
\[
I:\ x_0 \band \bnot x_1 \band x \band (y_0 = \bnot y_1) \band y \band z~,
\]
and transition relation
\[
T:\
\left[\begin{array}{l}
(x_0' = \bnot x_0) \band (x_1' = \bnot x_1) \band (x' = x_0 \bor x_1) \\
\mbox{} \band 
(y_0' = x \band \bnot y_0) \band (y_1' = x \band \bnot y_1) \band (y' = y_0 \bor y_1) \\
\mbox{} \band 
(z' = x \band y)
\end{array}\right]~.
\]
The intention is that $x$ and $y$ --- and thus $z$ --- are always
$\vtrue$.  This intention is asserted as the safety assertion $P: z$.
We apply the algorithm to this transition system to prove the
invariance of $P$.

\begin{enumerate}
\item $F_{0}$ is initialized to $I$, each of $F_{1}, F_{2}, F_{3}, \ldots$ to $P$, and $k$ to $1$.

\item $F_{1} \band T \band \bnot P'$ is satisfiable.  One satisfying
  assignment yields the $\bnot P$-predecessor $s_1: \bnot x_0 \band
  \bnot x_1 \band \bnot x \band \bnot y_0 \band \bnot y_1 \band \bnot
  y \band z$.  Is $\bnot s_1$ inductive relative to $F_{1}$?  Yes, as
  $F_{1} \band \bnot s_1 \band T$ implies $\bnot s_1'$.  Inductive
  generalization of $s_1$ relative to $F_{1}$ yields the clause $c_1:
  x_0 \bor x$, where (1) $c_1 \subset \bnot s_1$, and (2) $c_1$ is
  inductive relative to $F_{1}$.  As Table \ref{table:illustration}
  illustrates, $c_1$ is conjoined at both levels 1 and 2 while still
  maintaining the invariants on the sequence $F_0,F_1,F_2,\ldots$
  discussed above.  The clause $c_1$ not only excludes $s_1$ but also
  many other states, which is the purpose of inductive generalization.

\begin{table}[tb]
\caption{Incremental construction of an inductive strengthening assertion}
\label{table:illustration}
\centering
\vspace{-1ex}
$\begin{array}{|c|c|c|c|c|c|c|c|c|}
\hline
\mbox{Level} & 0 & 1 & 2 & 3 & 4 & 5 & 6 & 7 \\ \hline\hline
F_{0} & I &     &     &     &     &     &     &     \\ \hline
F_{1} & P & c_1 & c_2 & c_3 & c_4 & c_5 & c_4 & c_6 \\ \hline
F_{2} & P & c_1 & c_2 &     &     & c_5 & c_4 & c_6 \\ \hline
\end{array}\quad\quad\quad
\begin{array}{l@{\quad}l}
c_1: x_0 \bor x &       c_4: x_0 \bor x_1 \\
c_2: x_1 \bor x &       c_5: \bnot x_0 \bor \bnot x_1 \\
c_3: \bnot y_0 \bor y & c_6: x
\end{array}
$
\vspace{1em}
\end{table}

\item $F_{1} \band T \band \bnot P'$ is still satisfiable.  One
  satisfying assignment yields the $\bnot P$-predecessor $s_2: x_0
  \band \bnot x_1 \band \bnot x \band \bnot y_0 \band \bnot y_1 \band
  \bnot y \band z$.  $\bnot s_2$ is inductive relative to $F_{1}$.
  Inductive generalization yields from $\bnot s_2$ the clause $c_2:
  x_1 \bor x$, which is also inductive relative to $F_{1}$.

\item $F_{1} \band T \band \bnot P'$ is still satisfiable.  One
  satisfying assignment yields the $\bnot P$-predecessor $s_3: x_0
  \band x_1 \band \bnot x \band y_0 \band y_1 \band \bnot y \band z$,
  which has predecessor $s_4: \bnot x_0 \band \bnot x_1 \band x \band
  \bnot y_0 \band \bnot y_1 \band y \band z$ at level 1.  Hence,
  $\bnot s_3$ is not inductive relative to $F_{1}$.  However, it is
  inductive relative to $F_{0}$, and inductive generalization yields
  from $\bnot s_3$ the clause $c_3: \bnot y_0 \bor y$ at level 0.  As
  Table \ref{table:illustration} indicates, $c_3$ is only placed at
  level 1 (and implicitly at level 0).

\item The state $s_3$ is again considered at level 1, but as $c_3$
  does not exclude $s_4$, $\bnot s_3$ is still not inductive relative
  to $F_{1}$.  Therefore $s_4$ is considered.  But it, too, has a
  predecessor $s_5: x_0 \band x_1 \band x \band y_0 \band y_1 \band y
  \band z$ at level 1.  However, it is inductive relative to $F_{0}$,
  and inductive generalization yields $c_4: x_0 \bor x_1$ at level 0.

\item Now either $s_3$ or $s_4$ must be considered at level 1.
  Choosing $s_3$ reveals that $\bnot s_3$ is now inductive relative to
  $F_{1}$, and inductive generalization yields $c_5: \bnot x_0 \bor
  \bnot x_1$ at level 1.  Notice how the deduction of $c_4$ at level 0
  is crucial to the deduction of $c_5$ at level 1.

\item To finish this iteration, it remains to address $s_4$ at level
  1.  With the addition of $c_5$, $\bnot s_4$ is inductive relative to
  $F_{1}$, and inductive generalization yields again the clause $c_4:
  x_0 \bor x_1$, but now at level 1 instead of level 0.  Inductively
  generalizing cubes at the highest possible levels until convergence
  at $k$ makes it possible to deduce the equivalence $x_0 = \bnot
  x_1$, which requires two clauses to express.

\item $F_{1} \band T \band \bnot P'$ is still satisfiable.  One
  satisfying assignment yields the $\bnot P$-predecessor $s_6: x_0
  \band \bnot x_1 \band \bnot x \band y_0 \band y_1 \band \bnot y
  \band z$, which is inductive relative to $F_{1}$.  Inductive
  generalization yields the clause $c_6: x$ at level 1.

\item With $x$ at level 1, analysis of the $y$ component of the
  transition system proceeds similarly until $F_{1} \band T \band \bnot
  P'$ becomes unsatisfiable.

\item Propagation from $F_{1}$ to $F_{2}$ and from $F_{2}$ to $F_{3}$
  reveals that all clauses are inductive and inductively strengthen
  $z$.  Simplifying through subsumption and rewriting the formula
  yields the expected inductive strengthening
\[
x_0 = \bnot x_1 \band x \band y_0 = \bnot y_1 \band y \band z
\]
of the safety assertion $P: z$, thus proving its invariance.
\end{enumerate}

\subsection{Formal Presentation and Analysis}
\label{subsec:formal}

We present the algorithm and its proof of correctness simultaneously
with formally annotated pseudocode in Listings
\ref{code:prove}-\ref{code:push} using the classic approach to program
verification \cite{Floyd:Verification:1967,Hoare:Verification:1969}.
All assertions are inductive, but the ranking functions require some
additional reasoning.  For convenience, some assertions are labeled
and subsequently referenced in annotations.

\newcommand{\vmin}{\mathit{min}}
\newcommand{\vrv}{\mathit{rv}}
\newcommand{\vstates}{\mathit{states}}
\lstset{
  mathescape=true,
  basicstyle={\small\ttfamily},
  numbers=right,
  numberstyle={\scriptsize\sffamily},
  numbersep=3pt,
  firstnumber=auto,
  frame=single,
  framesep=2pt
}
\begin{lstlisting}[name=code,caption={The main function},label=code:prove,float=tb]
-post: $\vrv$ iff $P$ is invariant
bool prove():
  if either $I \band \bnot P$ or $I \band T \band \bnot P'$ is satisfiable:
    -assert: there exists a counterexample trace
    return false
  $F_{0}$ := $I$, clauses($F_0$) := $\emptyset$
  $F_{i}$ := $P$, clauses($F_i$) := $\emptyset$ for all $i > 0$
  for $k$ = $1$ to $\ldots$:
    -rank: at most $2^{|\vx|}+1$
    -assert ($A$): 
      (1) $\forall\ i \ge 0,\ I \Bimplies F_{i}$
      (2) $\forall\ i \ge 0,\ F_{i} \Bimplies P$
      (3) $\forall\ i > 0,\ \mathtt{clauses}(F_{i+1}) \subseteq \mathtt{clauses}(F_{i})$
      (4) $\forall\ 0 \le i < k,\ F_{i} \band T \Bimplies F_{i+1}'$
      (5) $\forall\ i > k,\ |\mbox{clauses}(F_i)| = 0$
    if not check($k$):
      -assert: there exists a counterexample trace
      return false
    propagate($k$)
    if there exists $1 \le i \le k$ such that $\mbox{clauses}(F_{i}) = \mbox{clauses}(F_{i+1})$:
      -assert:
        (1) $I \Bimplies F_{i}$
        (2) $F_{i} \band T \Bimplies F_{i}'$
        (3) $F_{i} \Bimplies P$
      return true
\end{lstlisting}

Listing \ref{code:prove} presents the top-level function {\tt prove},
which returns $\vtrue$ if and only if $P$ is invariant.  First it
looks for 0-step and 1-step counterexample traces.  If none are found,
$F_0,F_1,F_2,\ldots$ are initialized to assume that $P$ is invariant,
while their clause sets are initialized to empty.  As a formula, $F_i$
for $i > 0$ is interpreted as $P \band \bigwedge \mbox{\tt
  clauses}(F_i)$.  Then it constructs the sequence of $k$-step
over-approximations starting with $k=1$.  On each iteration, it first
calls {\tt check}($k$) (Listing \ref{code:check}), which strengthens
$F_{i}$ for $1 \le i \le k$ so that $F_{i}$-states are at least
$k-i+1$ steps away from violating $P$.  Then it calls {\tt
  propagate}($k$) (Listing \ref{code:check}) to propagate clauses
forward through $F_{1},F_{2},\ldots,F_{k+1}$ based on their having
become inductive relative to higher levels during the call to {\tt
  check}.  If this propagation yields any adjacent levels that share
all clauses (a simple syntactic check, not a validity check), an
inductive strengthening of $P$ has been discovered.

While the assertions are inductive, an argument needs to be made to
justify the ranking function.  By $A$.{\tt 3}, the state sets
represented by $F_{0},F_{1},\ldots,F_{k}$ are nondecreasing with
level.  To avoid termination at the {\tt if} check requires that they
be strictly increasing with level, which is impossible when $k$
exceeds the number of possible states.  Hence, $k$ is bounded by
$2^{|\vx|}+1$, and, assuming that the called functions always
terminate, {\tt prove} always terminates.

\begin{lstlisting}[name=code,caption={The {\tt check} and {\tt propagate} functions},label=code:check,float=tb]
-pre:  
  (1) $A$
  (2) $k \ge 1$
-post: 
  (1) $A$.1-3
  (2) if $\vrv$ then $\forall\ 0 \le i \le k,\ F_{i} \band T \Bimplies F_{i+1}'$
  (3) $\forall\ i > k+1,\ |\mbox{clauses}(F_i)| = 0$
  (4) if not $\vrv$ then there exists a counterexample trace
bool check($k$ : level):
  try:
    while $F_{k} \band T \band \bnot P'$ is satisfiable:
      -rank: at most $2^{|\vx|}$
      -assert ($B$):
        (1) $A$.1-4
        (2) $\forall\ c \in \mbox{clauses}(F_{k+1}),\ F_{k} \band T \Bimplies c'$
        (3) $\forall\ i > k+1,\ |\mbox{clauses}(F_i)| = 0$
      let $s$ be the predecessor extracted from the witness
      -assert: $k < 2$ or $\bnot s$ is inductive relative to $F_{k-2}$
      $n$ := inductive($s$, $k-2$, $k$)
      push($\set{(n+1, s)}$, $k$)
      -assert ($C$): $\notmodels{F_{k}}{s}$
    return true
  except Counterexample:
    return false

-pre/post:
  (1) $A$.1-3
  (2) $\forall\ 0 \le i \le k,\ F_{i} \band T \Bimplies F_{i+1}'$
  (3) $\forall\ i > k+1,\ |\mbox{clauses}(F_i)| = 0$
void propagate($k$ : level):
  for $i$ = $1$ to $k$:
    for each $c$ in clauses($F_{i}$):
      -assert: pre/post
      if $F_{i} \band T \band \bnot c'$ is unsatisfiable:
        $F_{i+1}$ := $F_{i+1} \band c$
\end{lstlisting}

For a given level $k$, {\tt check}($k$) (Listing \ref{code:check})
iterates until $F_{k}$ excludes all states that can lead to a violation
of $P$ in one step.  Suppose $s$ is one such state.  It is eliminated
by, first, inductively generalizing it at the highest level $n$ at
which $\bnot s$ is inductive relative to $F_{n}$ through a call to {\tt
  inductive}($s$, $k-2$, $k$) (Listing \ref{code:ig}) and then,
second, pushing for a generalization at level $k$ through a call to
{\tt push}($\set{(n+1, s)}$, $k$) (Listing \ref{code:push}).  At the
end of the iteration, $F_{k}$ excludes $s$ (assertion $C$).  This
progress implies that the loop can iterate at most as many times as
there are possible states, yielding {\tt check}'s ranking function.

Notice how {\tt check}, according to its postcondition, preserves loop
invariants $A$.{\tt 1-3} while incrementing $A$.{\tt 4-5} to to apply
to an additional step (see postconditions (2) and (3)), unless a
counterexample is found.

\begin{lstlisting}[name=code,caption={$i$-step relative inductive generalization},label=code:ig,float=tb]
-pre:  
  (1) $B$
  (2) $i \ge 0$
  (3) $\bnot s$ is inductive relative to $F_{i}$
-post: 
  (1) $B$
  (2) $\notmodels{F_{i+1}}{s}$
void generate($s$ : state, $i$ : level, $k$ : level):
  $c$ := find subclause of $\bnot s$ that is inductive relative to $F_{i}$
  for $j$ = $1$ to $i+1$:
    -assert: 
      (1) $B$
      (2) $\notmodels{F_{j-1}}{s}$
    $F_{j}$ := $F_{j} \band c$

-pre:  
  (1) $B$
  (2) $\vmin \ge -1$
  (3) $\vmin < 0$ or $\bnot s$ is inductive relative to $F_{\vmin}$
  (4) there is a trace from $s$ to a $\bnot P$-state
-post: 
  (1) $B$
  (2) $\vmin \le rv \le k$, $rv \ge 0$
  (3) $\notmodels{F_{rv+1}}{s}$
  (4) $\bnot s$ is inductive relative to $F_{rv}$
level inductive($s$ : state, $\vmin$ : level, $k$ : level):
  if $\vmin < 0$ and $F_{0} \band T \band \bnot s \band s'$ is satisfiable:
    -assert: there exists a counterexample trace
    raise Counterexample
  for $i$ = max(1, $\vmin+1$) to $k$:
    -assert: 
      (1) $B$
      (2) $\vmin < i \le k$
      (3) $\forall\ 0 \le j < i$, $\bnot s$ is inductive relative to $F_{j}$
    if $F_{i} \band T \band \bnot s \band s'$ is satisfiable:
      generate($s$, $i-1$, $k$)
      return $i-1$
  generate($s$, $k$, $k$)
  return $k$
\end{lstlisting}

The functions {\tt inductive} and {\tt generate} (Listing
\ref{code:ig}) perform inductive generalization.  The details of
discovering an inductive subclause are described in previous work
\cite{Bradley+Manna/2007}.  One interesting observation, however, is
that when calling {\tt inductive}, a minimum level $\vmin$ at which
$\bnot s$ is inductive relative to $F_{\vmin}$ can be supplied.  At
lines 43-44, $\notmodels{F_{k-1}}{s}$ by $A$.{\tt 2} and $A$.{\tt 4}
so that $\bnot s$ is inductive relative to $F_{k-2}$ by $A$.{\tt 4}.
At lines 127-128, $\bnot s$ is inductive relative to $F_{n-1}$ so that
$\notmodels{F_{n-1}}{p}$ and thus $\bnot p$ is inductive relative to
$F_{n-2}$ by $A$.{\tt 4}.  If $\vmin < 0$, then it is possible that
$s$ is reachable from an initial state, hence the check at line 87.

\begin{lstlisting}[name=code,caption={The {\tt push} function for $k$-step relative inductive generalization},label=code:push,float=tb]
-pre:  
  (1) $B$
  (2) $\forall\ (i, q) \in \vstates,\ 0 < i \le k+1$
  (3) $\forall\ (i, q) \in \vstates,\ \notmodels{F_i}{q}$
  (4) $\forall\ (i, q) \in \vstates,\ \bnot q$ is inductive relative to $F_{i-1}$
  (5) $\forall\ (i, q) \in \vstates$, there is a trace from $q$ to a $\bnot P$-state
-post:
  (1) $B$
  (2) $\forall\ (i, q) \in \vstates,\ \notmodels{F_{k}}{q}$
void push($\vstates$ : (level, state) set, $k$ : level):
  while true:
    -rank: at most $(k+1) 2^{|\vx|}$
    -assert ($D$):
      (1) $B$
      (2) $\forall\ (i, q) \in \vstates_{\mathsf{prev}},\ \exists j \ge i,\ (j, q) \in \vstates$
      (3) $\forall\ (i, q) \in \vstates,\ 0 < i \le k+1$
      (4) $\forall\ (i, q) \in \vstates,\ \notmodels{F_{i}}{q}$
      (5) $\forall\ (i, q) \in \vstates,\ \bnot q$ is inductive relative to $F_{i-1}$
      (6) $\forall\ (i, q) \in \vstates,$ there is a trace from $q$ to a $\bnot P$-state
    ($n$, $s$) := choose pair from $\vstates$ that minimizes $n$
    -assert: $\forall\ (i, q) \in \vstates,\ n \le i$
    if $n > k$:
      return
    if $F_{n} \band T \band s'$ is satisfiable:
      let $p$ be the predecessor extracted from the witness
      -assert ($E$):
        (1) $\forall\ (i, q) \in \vstates,\ p \ne q$
        (2) $n < 2$ or $\bnot p$ is inductive relative to $F_{n-2}$
      $m$ := inductive($p$, $n-2$, $k$)
      $\vstates$ := $\vstates \cup \set{(m+1, p)}$
    else:
      $m$ := inductive($s$, $n$, $k$)
      -assert ($F$): $m+1 > n$
      $\vstates$ := $\vstates \setminus \set{(n, s)} \cup \set{(m+1, s)}$
\end{lstlisting}

The {\tt push} algorithm (Listing \ref{code:push}) is the key to
``pushing'' inductive generalization to higher levels.  The insight is
simple: if a state $s$ is not inductive relative to $F_{i}$, apply
inductive generalization to its predecessors that satisfy $F_{i}$.
The complication is that this recursive analysis must proceed in a
manner that terminates despite the presence of cycles in the system's
state graph.  To achieve termination, a set $\vstates$ of pairs $(i,
s)$ is maintained such that each pair $(i, s) \in \vstates$ represents
the knowledge that (1) $s$ is inductive relative to $F_{i-1}$, and (2)
$F_{i}$ excludes $s$.  The loop in {\tt push} always selects a pair
$(n, s)$ from $\vstates$ such that $n$ is minimal over the set.
Hence, none of the states already represented in $\vstates$ can be a
predecessor of $s$ at level $n$.

Formally, termination of {\tt push} is established by the inductive
assertions $D$.{\tt 2}, which asserts that the set of states
represented in $\vstates$ does not decrease; $E$.{\tt 1}, which
asserts that each state in $\vstates$ is represented by at most one
pair in $\vstates$; and $F$, which asserts that the level associated
with a state can only increase.  Given that each iteration either adds
a new state to $\vstates$ or increases a level for some state already
in $\vstates$ and that levels peak at $k+1$, the number of iterations
is bounded by the product of $k+1$ and the size of the state space.

The inductive proof in Listings \ref{code:prove}-\ref{code:push} and
the termination arguments yield total correctness:

\begin{theorem} 
\label{th:correct}
For finite transition system $S: (\vx, I, T)$, the algorithm always
terminates and returns true if and only if safety assertion $P$ is
invariant.
\end{theorem}

\subsection{Variations}

Notice that {\tt inductive} and {\tt generate} (Listing \ref{code:ig})
together generate a subclause of $\bnot s$ that is inductive relative
to $F_{i}$, where $i$ is the greatest level for which $\bnot s$ is
itself inductive relative to $F_{i}$.  It is actually possible to find
the highest level $j \ge i$ for which $\bnot s$ has a subclause that
is inductive relative to $F_{j}$ even if $\bnot s$ is not itself
inductive relative to $F_{j}$ (that is, $j > i$).  The difference
between these two approaches is in whether the {\sf down} function of
\cite{Bradley+Manna/2007} is ever applied to $\bnot s$.  In the method
of {\tt inductive} and {\tt generate}, it is not; in the variation, it
is.

While generalizing at higher levels is desirable, applying {\sf down}
to large clauses, such as $\bnot s$, is the most expensive phase of
inductive generalization in practice.  On particularly large
benchmarks with thousands of latches this phase can take prohibitively
long; for example, on the {\tt neclaftpX00X} benchmarks from HWMCC'08,
this variation does not typically terminate in under 15 minutes.

One might wonder, therefore, if a weaker but faster inductive
generalization procedure could be used.  An obvious such procedure is
the following: rather than using full induction, one could search for
clauses that are established in the next state without assuming them
as inductive hypotheses --- in other words, perform a search for an
implicate subclause (that is also inductive) rather than for an
inductive subclause.  Experiments indicate that using this
generalization yields an overall model checker that is rarely faster
and often significantly slower despite the superior speed of the
individual generalizations.  Of course, a positive spin on this
disappointing result is that full induction is apparently a powerful
generalization technique compared to searching for implicates.

\section{Implementation and Experiments}
\label{sec:exp}

\subsection{Implementation}

We implemented the algorithm using O'Caml for top-level reasoning,
MiniSAT 2.0 for preprocessing the transition relation
\cite{Een+Biere/2005}, and ZChaff for SAT-solving because of its
incremental solving capability \cite{Moskewicz+Others/2001}.  Notice
that the SAT-solving libraries were available before 2008; thus, our
performance on the HWMCC'08 benchmarks reported below cannot be
attributed to superior SAT solvers.

{\bf Preprocessing.}  MiniSAT 2.0 provides an interface for
``freezing'' variables that should not be chosen for elimination
during preprocessing.  We use it to simplify the given transition
relation once and for all \cite{Een+Others/2007}.  Reducing the
transition relation according to the cone-of-influence
\cite{Clarke+Others:MC:2000} followed by preprocessing yielded
significant performance improvements for inductive generalization.  It
is likely that more sophisticated preprocessing would yield better
performance.

{\bf Incremental SAT-Solving.}  Our technique requires solving
hundreds to thousands of SAT problems per second in an incremental
fashion.  While MiniSAT 2.0 provides the ability to maintain context
and change assumptions in the form of literals, only ZChaff, as far as
we know, provides competitive SAT-solving combined with the ability to
push and pop incremental context that includes sets of clauses.  It is
likely that a fully incremental version of a modern SAT solver would
yield better performance.

{\bf Optimizations.}  Given that our algorithm relies on inductive
generalization, we implemented a simple method to extract literal
invariants that are obvious from the structure of the initial
condition and transition relation.  This optimization greatly improved
performance on the {\tt neclaftpX00X} benchmarks.

We implemented binary, rather than linear, search in the function {\tt
  inductive}.

In our implementation of inductive generalization
\cite{Bradley+Manna/2007}, we use a simple threshold to end the search
for a minimal inductive subclause.  If a certain number of randomly
chosen literals (three in our implementation) are determined to be
necessary to yield an inductive subclause, the search for a smaller
inductive subclause ends.  While minimality is no longer guaranteed,
the resulting clauses are sufficiently strong (and probably minimal).

Finally, we implemented a VSIDS-like literal-ordering heuristic to
guide which inductive clauses are discovered from a given cube
\cite{Moskewicz+Others/2001}.  Since a given clause can have many
minimal inductive subclauses, the idea is to focus on those literals
whose negations have appeared most frequently in examined states in
recent history.  Unfortunately, whether the heuristic has any benefit
is unclear.



\subsection{Experiments}

The benchmarks and results from the Hardware Model Checking
Competition 2008 provide a means of comparing different model checking
algorithms \cite{hwmcc08}.  We report our performance on these
benchmarks.

We performed all experiments on a laptop equipped with an Intel Core 2
Duo 2.2 GHz processor, although only one core was used, and 4 GB of
memory.  In the HWMCC'08 competition, entries ran on Pentium IV 3 GHz
processors with 2 GB of memory.  After reading various online forums,
we concluded that our processor provides a speed advantage of
approximately $1.8\times$ over the hardware used in the competition.
Thus, rather than counting the number of benchmarks solved in under
900 seconds, we count only those solved in under 500 seconds.

Our implementation constructs proofs of unsatisfiability for 325
benchmarks in under 500 seconds and using at most 1.5 GB of memory,
compared to the 314 solved by {\tt abc}, the winner of the {\it unsat}
division of the competition.  Ten of these benchmarks were not solved
during the competition.  It finds counterexample traces in 234 cases,
surprisingly competitive with BMC \cite{Biere+Others/1999}.  The top
four entries for the satisfiable problems, all based on BMC, found
247, 243, 239, and 239 counterexamples, respectively.  Our total
number of solved problems is thus 559, seven more than {\tt abc}, the
winner of the overall competition.

Table \ref{table:data} presents data for the 38 benchmarks that our
implementation proved unsatisfiable in the allotted time (500 seconds)
and memory (1.5 GB) that were solved by at most three competitors.  The
second column lists those competitors who solved the benchmark, their
time in seconds (unscaled), and their peak memory consumption in MB.
The third through sixth columns list our implementation's time in
seconds {\it scaled by 1.8 to allow for better comparison}, memory
consumption in MB, the number of thousands of SAT instances solved,
and the number of the clauses in the proof, respectively.  Again, the
time for our implementation is multiplied by 1.8, so indicated runtime
can be over 500 seconds despite our setting the timeout at 500
seconds.

\begin{table}[tb]
\label{table:data}
\caption{Solved benchmarks that were solved in HWMCC'08 by at most three solvers}
\centering
\vspace{-1ex}
{\scriptsize
\begin{tabular}{|l|l|r|r|r|r|}
\hline
{\bf Benchmark} & {\bf Solved by} ({\sf solver}/sec/MB) & {\bf Sec} & {\bf MB} & {\bf SC}(k) & {\bf $|$Proof$|$} \\ \hline\hline
\input{data}
\end{tabular}
\vspace{1em}
}
\end{table}

In case the 1.8 scaling to compensate for different processors is
considered too low, the results for 3.0 scaling are the following: 317
proofs and 228 counterexamples, with 545 benchmarks solved overall.

\section{Related Work}
\label{sec:related}


SAT-based unbounded model checking was the first symbolic model
checking approach based on generating clauses \cite{McMillan/2002}.
It discovers implicates to generalize states leading to property
violations.  The overall iterative structure is the same as standard
symbolic model checking.  In our algorithm, induction is a means not
only for generalizing from states but also for abstracting the system
based on the property, allowing the analysis of large transition
systems.


Our algorithm can be seen as an instance of predicate
abstraction/refinement \cite{Graf+Saidi/1997,Clarke+Others/2003} in
that the minor iterations generate new predicates (clauses) while the
major iterations propagate them.  If the clauses are insufficient for
convergence to an inductive strengthening assertion, the next minor
iteration generates additional clauses that allow propagation to
continue at least one additional step.


The $k$-step over-approximation structure of $F_{0}, F_{1}, F_{2},
\ldots, F_{k}$ is similar to that of interpolation-based model checking
(ITP) \cite{McMillan/2003}, which uses an interpolant from an
unsatisfiable $K$-step BMC query to compute the post-image
approximately.  All states in the image are at least $K-1$ steps away
from violating the property.  A larger $K$ refines the image by
increasing the minimum distance to violating states.  In our
algorithm, if the frontier is at level $k$, then $F_{i}$, for $0 \le i
\le k$, represents states that are at least $k-i$ steps from violating
the property.  As $k$ increases, the minimum number of steps from
$F_{i}$-states to violating states increases.  In both cases,
increasing $k$ (in ours) or $K$ (in ITP) sufficiently for a correct
system yields an inductive assertion.  However, the algorithms differ
in their underlying ``technology'': ITP computes interpolants from
$K$-step BMC queries, while our algorithm uses inductive
generalization of cubes, which requires only 1-step BMC queries for
arbitrarily large $k$.


Various approaches to generalizing counterexamples to $k$-induction
have been explored
\cite{DeMoura+Others/2003,Awedh+Others/2006,Vimjam+Others/2006}.  Our
work could in principle be applied as a method of strengthening
$k$-induction.  However, the technique already works well on its own
and has the distinct advantage of posing small SAT problems.

Finally, we draw on our previous work on inductive generalization
\cite{Bradley+Manna/2007}.  This paper contributes $k$-step relative
inductive generalization, which guarantees that all examined cubes can
be inductively generalized if the property is invariant.

\section{Conclusion}

The empirical data suggest the effectiveness of $k$-step relative
inductive generalization, a technique unlike --- and therefore
complementary to --- other symbolic model checking methods.  The most
exciting direction for our ongoing research is to parallelize the
algorithm.  Our earlier work on inductive generalization was easily
parallelized and sometimes yielded near-linear scaling with the number
of nodes on hard benchmarks \cite{Bradley/2007}.  The new algorithm,
although more complex in structure, should be similarly parallelizable
since the implementation spends the majority of its time generating
clauses incrementally.

BMC is faster than our implementation at finding counterexample
traces.  We plan to investigate a combination of our algorithm with
BMC in which generated clauses would constrain the SAT search space.

Another direction for research is to apply the idea of finding
$k$-step relative inductive generalizations of states in an
infinite-state setting.

\bibliographystyle{acm}

\end{document}

%% file: data.tex
{\tt bjrb07amba6andenv} & {\sf abc}/309/166 {\sf pdtravbdd}/19/61 & 462 & 364 & 11 & 269 \\ \hline
{\tt bjrb07amba7andenv} & {\sf abc}/203/180 {\sf pdtravbdd}/242/71 & 169 & 253 & 7 & 221 \\ \hline
{\tt intel006} & {\sf pdtravitp}/348/143 {\sf tipidi}/367/425 & 32 & 79 & 28 & 931 \\ \hline
{\tt intel007} & {\sf pdtravcbq}/881/185 & 541 & 228 & 76 & 2906 \\ \hline
{\tt intel026} & & 261 & 277 & 96 & 1335 \\ \hline
{\tt intel037} & & 207 & 786 & 2 & 157 \\ \hline
{\tt intel054} & {\sf tipidi}/2/8 {\sf tipids}/2/8 {\sf tipind}/2/8 & 414 & 174 & 271 & 4544 \\ \hline
{\tt intel055} & {\sf tipidi}/43/12 {\sf tipids}/43/12 & 39 & 95 & 30 & 615 \\ \hline
{\tt intel056} & {\sf tipidi}/7/13 {\sf tipids}/8/13 & 91 & 79 & 93 & 1597 \\ \hline
{\tt intel057} & {\sf tipidi}/2/6 {\sf tipids}/2/6 {\sf tipind}/2/6 & 176 & 129 & 142 & 2332 \\ \hline
{\tt intel059} & {\sf tipidi}/4/8 {\sf tipids}/4/8 & 46 & 74 & 53 & 982 \\ \hline
{\tt neclabakery001} & {\sf aigtrav}/14/95 {\sf pdtravbdd}/18/54 {\sf tipind}/422/34 & 156 & 233 & 417 & 2755 \\ \hline
{\tt neclaftp1001} & & 84 & 781 & 1 & 669 \\ \hline
{\tt neclaftp1002} & & 284 & 1417 & 3 & 707 \\ \hline
{\tt neclaftp2001} & {\sf tipidi}/839/122 {\sf tipids}/838/122 {\sf tipind}/834/123 & 43 & 466 & 1 & 638 \\ \hline
{\tt neclaftp2002} & {\sf tipind}/898/175 & 248 & 816 & 3 & 644 \\ \hline
{\tt neclatcas1a001} & {\sf tipidi}/0/0 {\sf tipids}/0/0 {\sf tipind}/0/0 & 3 & 56 & 1 & 86 \\ \hline
{\tt neclatcasall001} & {\sf tipidi}/0/0 {\sf tipids}/0/0 {\sf tipind}/0/0 & 45 & 97 & 20 & 279 \\ \hline
{\tt nusmvbrp} & {\sf pdtravbdd}/456/74 {\sf pdtravcbq}/187/283 & 21 & 50 & 56 & 688 \\ \hline
{\tt nusmvguidancep2} & {\sf pdtravbdd}/478/61 {\sf tipidi}/873/394 & 23 & 78 & 16 & 164 \\ \hline
{\tt nusmvguidancep5} & {\sf pdtravbdd}/59/44 & 16 & 70 & 10 & 121 \\ \hline
{\tt nusmvguidancep6} & {\sf abc}/34/35 {\sf pdtravbdd}/54/44 {\sf pdtravitp}/92/177 & 10 & 69 & 8 & 97 \\ \hline
{\tt pdtvisbakery0} & {\sf abc}/21/97 {\sf pdtravbdd}/28/60 & 113 & 164 & 36 & 215 \\ \hline
{\tt pdtvisbakery1} & {\sf abc}/96/97 {\sf pdtravbdd}/44/61 & 144 & 182 & 44 & 308 \\ \hline
{\tt pdtvisbakery2} & {\sf abc}/57/95 {\sf pdtravbdd}/114/64 & 136 & 191 & 42 & 371 \\ \hline
{\tt pdtvisgoodbakery0} & {\sf abc}/45/98 {\sf pdtravbdd}/57/64 & 203 & 202 & 65 & 601 \\ \hline
{\tt pdtvisgoodbakery1} & {\sf abc}/102/95 {\sf pdtravbdd}/51/63 & 142 & 175 & 46 & 458 \\ \hline
{\tt pdtvisgoodbakery2} & {\sf abc}/118/97 {\sf pdtravbdd}/49/60 & 153 & 193 & 47 & 372 \\ \hline
{\tt pdtvisns3p00} & & 244 & 138 & 120 & 1709 \\ \hline
{\tt pdtvisns3p01} & {\sf pdtravcbq}/618/266 {\sf tipids}/670/145 & 352 & 131 & 152 & 2287 \\ \hline
{\tt pdtvisns3p02} & & 196 & 129 & 100 & 1169 \\ \hline
{\tt pdtvisns3p03} & & 230 & 121 & 106 & 1398 \\ \hline
{\tt pdtvisns3p04} & & 550 & 115 & 207 & 2187 \\ \hline
{\tt pdtvisns3p06} & {\sf pdtravcbq}/823/278 & 837 & 164 & 289 & 2845 \\ \hline
{\tt pdtvisns3p07} & & 311 & 131 & 145 & 1453 \\ \hline
{\tt pdtvisrethersqo4} & {\sf abc}/23/15 & 162 & 157 & 341 & 3394 \\ \hline
{\tt pdtvissoap1} & {\sf pdtravitp}/384/520 & 70 & 102 & 42 & 807 \\ \hline
{\tt pdtvissoap2} & & 108 & 101 & 65 & 1789 \\ \hline